\documentclass[12pt,epsf,amssymb,qsymbols]{article}
\usepackage{tabularx}
\usepackage{array}
\usepackage{graphics}
\usepackage{graphicx}
\usepackage{psfrag}
\usepackage{epsfig}
\usepackage{amsmath}
\usepackage{amssymb}
\usepackage{ulem}
\usepackage{setspace}
\usepackage{rotating}
\usepackage{colortbl}
\usepackage{tabularx}
\usepackage{longtable}
\usepackage{multirow}
\makeatletter

\usepackage{verbatim}

\setlongtables

\newcommand{\msbar}{{\overline{\rm MS}}}

\newcommand{\bea}{\begin{eqnarray}}
\newcommand{\eea}{\end{eqnarray}}
\newcommand{\beq}{\begin{equation}}
\newcommand{\eeq}{\end{equation}}
\newcommand{\ec}{\end{center}}
\newcommand{\bc}{\begin{center}}
\newcommand{\gev}{{\rm GeV}}

\newcommand{\pdir}{p\kern -5.2pt\raise 0.2ex\hbox {/}}

\newcommand{\vdir}{v\kern -5.75pt\raise 0.15ex\hbox {/}}
\newcommand{\kdir}{k\kern -5.75pt\raise 0.15ex\hbox {/}}
\newcommand{\epsdir}{\epsilon\kern -5.0pt\raise 0.15ex\hbox {/}}
\newcommand{\bvdir}{\bar{v}\kern -5.75pt\raise 0.15ex\hbox {/}}
\newcommand{\Ddir}{D\kern -7.75pt\raise 0.20ex\hbox {/}}
\newcommand{\Adir}{A\kern -7.75pt\raise 0.20ex\hbox {/}}
\newcommand{\ldir}{l\kern -5.0pt\raise 0.2ex\hbox{/}}
\newcommand{\varepsdir}{\varepsilon\kern -5.5pt\raise 0.15ex\hbox{/}}

\def \eff{{\text{eff}}}

\newcommand{\nn}{\nonumber}
\makeatother

\begin{document}
\thispagestyle{empty} 
\begin{flushright}
\begin{tabular}{l}
LPT 11-50
\end{tabular}
\end{flushright}
\begin{center}
\vskip 3.0cm\par
{\par\centering \textbf{\LARGE  
\Large \bf On transverse asymmetries in $B\to K^\ast \ell^+\ell^-$ }}\\
\vskip 1.25cm\par
{\scalebox{.87}{\par\centering \large  
\sc Damir Be\v{c}irevi\'c$^a$ and Elia Schneider$^{a,b}$ }}
{\par\centering \vskip 0.5 cm\par}
{\sl 
$^a$~Laboratoire de Physique Th\'eorique (B\^at.~210)~\footnote{Laboratoire de Physique Th\'eorique est une unit\'e mixte de recherche du CNRS, UMR 8627.}\\
Universit\'e Paris Sud, Centre d'Orsay,\\ 
F-91405 Orsay-Cedex, France.}\\
{\par\centering \vskip 0.25 cm\par}
{\sl 
$^b$~Dipartimento di Fisica ``EnricoFermiÓ, Universit\`a di Pisa, \\
Largo B.~Pontecorvo 3, I-56127 Pisa, Italy.}\\
 
{\vskip 1.35cm\par}
\end{center}

\vskip 0.55cm
\begin{abstract}
We discuss the three independent asymmetries, $A_T^{{\rm (2)}}(q^2)$, $A_T^{{\rm (im)}}(q^2)$ and  $A_T^{{\rm (re)}}(q^2)$, that one can build from the amplitudes $A_{\perp} (q^2)$ and  $A_{\parallel} (q^2)$. These quantities are expected to be accessible from the new $B$-physics experiments, they are sensitive to the presence of new physics, and they are not very sensitive to hadronic uncertainties. Studying their low $q^2$ dependence can be helpful in discerning among various possible new physics scenarios. All three asymmetries can be extracted from the full angular analysis of $B\to K^\ast \ell^+\ell^-$. Our formulas apply to both the massless and the massive lepton case. 
\end{abstract}
\vskip 2.6cm
{\small PACS: 12.20.He} 
\newpage
\setcounter{page}{1}
\setcounter{footnote}{0}
\setcounter{equation}{0}
\noindent

\renewcommand{\thefootnote}{\arabic{footnote}}

\setcounter{footnote}{0}
\section{Introduction and basic formulas}
Searches for physics beyond the Standard Model (BSM) through the low energy flavor physics experiments could provide us with an indirect tool for identifying the new particles that will hopefully be directly detected at  LHC. Flavor changing neutral currents are known to be particularly revealing in that respect, and the $b\to s$ transitions are known to be particularly interesting. In this paper we will focus on the exclusive $b\to s$ mode, $B\to K^\ast (K\pi) \ell^+\ell^-$, where $\ell$ stands for a lepton.~\footnote{In practice one considers either electron or muon, but not tau, although the expressions given in this paper apply equally to the $B\to K^\ast \tau^+\tau^-$ case.} After an intensive research devoted to this decay it has been understood that the full branching fraction is extremely  difficult to handle theoretically because: (1) all the hadronic form factors enter the corresponding expression, and (2) every form factor should be integrated over a very large range of kinematically accessible values of $q^2$, namely $q^2 \in [4m_\ell^2, q^2_{\rm max}]$, with $q^2_{\rm max} = (m_B-m_{K^\ast})^2 = 19.25\ \gev$. None of the available methods to compute form factors is viable in the entire physical range of momenta transferred to leptons, and therefore only the partial decay rates (integrated over a relatively short span of $q^2$'s) can be computed, albeit with a limited control over theoretical non-perturbative QCD uncertainties.  Furthermore, since the study of this decay is based on the use of operator product expansion (OPE), it is essential to avoid the resonances arising when the energy of the lepton pair $\ell^+\ell^-$ hits the production thresholds of the $c\bar c$-resonances (i.e. $J/\psi$ and its radial excitations). For that reason one should try and work below $q^2= m^2_{J/\psi}\approx 9.6\ \gev^2$, or experimentally veto the points at which the resonances are expected to occur. Besides, one can also encounter difficulties with the charmless resonances ($\rho,\omega$ and their radial excitations) but that effect turns out to be practically negligible thanks to the CKM suppression. 

Using OPE, at low energies, this decay is described by the following effective Hamiltonian~\cite{Heff}:
 \begin{equation} \label{eq:Heff}
  {\cal H}_{\eff} = - \frac{4\,G_F}{\sqrt{2}} V_{tb}V_{ts}^\ast
     \left[  \sum_{i=1}^{6} C_i (\mu)
\mathcal O_i(\mu) + \sum_{i=7,8,9,10,P,S} \biggl(C_i (\mu) \mathcal O_i + C'_i (\mu) \mathcal
O'_i\biggr)\right] \,,
\end{equation}
where the twice Cabibbo suppressed contribution ($\propto  V_{ub}V_{us}^\ast $) has been neglected. The operator basis looks as follows~\cite{Bobeth:1999mk,Altmannshofer:2008dz}:
\bea
{\mathcal{O}}_{1} &=&
(\bar{s} \gamma_{\mu} T^a P_L c)(\bar{c} \gamma^{\mu} T^a P_L b) ,  \nn \\
{\mathcal{O}}_{2} &=&
(\bar{s} \gamma_{\mu}  P_L c)(\bar{c} \gamma^{\mu}  P_L b) , \nn  \\
{\mathcal{O}}_{3} &=&
(\bar{s} \gamma_{\mu}  b) \sum_{q} (\bar{q} \gamma^{\mu}  q) , \nn \\
{\mathcal{O}}_{4} &=&
(\bar{s} \gamma_{\mu} T^a b) \sum_{q} (\bar{q} \gamma^{\mu}T^a  q) ,\nn   \\
{\mathcal{O}}_{5} &=&
(\bar{s} \gamma_{\mu_1} \gamma_{\mu_2}\gamma_{\mu_3} b) \sum_{q} (\bar{q} \gamma^{\mu_1}\gamma^{\mu_2}\gamma^{\mu_3}  q) ,\nn   \\
{\mathcal{O}}_{6} &=&
(\bar{s} \gamma_{\mu_1} \gamma_{\mu_2}\gamma_{\mu_3}T^a b) \sum_{q} (\bar{q} \gamma^{\mu_1}\gamma^{\mu_2}\gamma^{\mu_3}T^a  q) , \nn 
\eea  
\begin{align}
{\mathcal{O}}_{7} &= \frac{e}{g^2} m_b
(\bar{s} \sigma_{\mu \nu} P_R b) F^{\mu \nu} ,&
{\mathcal{O}}_{7}^\prime &= \frac{e}{g^2} m_b
(\bar{s} \sigma_{\mu \nu} P_L b) F^{\mu \nu} ,\nn \\
{\mathcal{O}}_{8} &= \frac{1}{g} m_b
(\bar{s} \sigma_{\mu \nu} T^a P_R b) G^{\mu \nu \, a} ,&
{\mathcal{O}}_{8}^\prime &= \frac{1}{g} m_b
(\bar{s} \sigma_{\mu \nu} T^a P_L b) G^{\mu \nu \, a} ,\nn \\
{\mathcal{O}}_{9} &= \frac{e^2}{g^2} 
(\bar{s} \gamma_{\mu} P_L b)(\bar{\ell} \gamma^\mu \ell) ,&
{\mathcal{O}}_{9}^\prime &= \frac{e^2}{g^2} 
(\bar{s} \gamma_{\mu} P_R b)(\bar{\ell} \gamma^\mu \ell) ,\nn \\
{\mathcal{O}}_{10} &=\frac{e^2}{g^2}
(\bar{s}  \gamma_{\mu} P_L b)(  \bar{\ell} \gamma^\mu \gamma_5 \ell) ,&
{\mathcal{O}}_{10}^\prime &=\frac{e^2}{g^2}
(\bar{s}  \gamma_{\mu} P_R b)(  \bar{\ell} \gamma^\mu \gamma_5 \ell) ,\nn \\
{\mathcal{O}}_{S} &=\frac{e^2}{16\pi^2}
m_b (\bar{s} P_R b)(  \bar{\ell} \ell) ,&
 {\mathcal{O}}_{S}^\prime &=\frac{e^2}{16\pi^2}
m_b (\bar{s} P_L b)(  \bar{\ell} \ell) ,\nn \\
{\mathcal{O}}_{P} &=\frac{e^2}{16\pi^2}
m_b (\bar{s} P_R b)(  \bar{\ell} \gamma_5 \ell) ,&
 {\mathcal{O}}_{P}^\prime &=\frac{e^2}{16\pi^2}
m_b (\bar{s} P_L b)(  \bar{\ell} \gamma_5 \ell),
\end{align}
with $P_{L,R}=(1\mp \gamma_5)/2$. The operators with the opposite chirality, $\mathcal O_{1-6}^\prime$, have been neglected. Short distance physics effects, encoded in the Wilson coefficients $C_i (\mu)$, have been computed in the Standard Model (SM) through a perturbative matching between the effective and full theories at $\mu=m_W$,  and then evolved down to the $\mu=m_b$ by means of the QCD renormalization group equations at next-to-next-to-leading logarithmic approximation (NNLO)~\cite{Bobeth:1999mk}. For the reader's convenience the resulting values of Wilson coefficients $C_i(m_b)$ are listed in Appendix of the present paper. 
An important feature regarding the operator $\mathcal O_9$ is its mixing with $\mathcal O_{1,\dots,6}$ through diagrams with a virtual photon decaying to
$\ell^+\ell^-$. It is customary to reassemble Wilson coefficients multiplying the same hadronic matrix element into effective coefficients appearing in the physical amplitudes, namely~\cite{Wilsoneff}
\bea
C_7^{\rm eff} & = & \frac{4\pi}{\alpha_s}\, C_7 -\frac{1}{3}\, C_3 -
\frac{4}{9}\, C_4 - \frac{20}{3}\, C_5\, -\frac{80}{9}\,C_6\,,
\nonumber\\
C_8^{\rm eff} & = & \frac{4\pi}{\alpha_s}\, C_8 + C_3 -
\frac{1}{6}\, C_4 + 20 C_5\, -\frac{10}{3}\,C_6\,,
\nonumber\\
C_9^{\rm eff} & = & \frac{4\pi}{\alpha_s}\,C_9 + Y(q^2)\,,
\nonumber\\
C_{10}^{\rm eff} & = & \frac{4\pi}{\alpha_s}\,C_{10}\,,\qquad
C_{7,8,9,10}^{\prime,\rm eff} = \frac{4\pi}{\alpha_s}\,C'_{7,8,9,10}\,,
\eea
where the function $Y(q^2)$ is
\bea\label{eq:Y}
Y(q^2) & = &  \frac{4}{3}\, C_3 + \frac{64}{9}\, C_5 + \frac{64}{27} C_6 -\frac{1}{2}\,h(q^2,0) \left( C_3 + \frac{4}{3}\,C_4 + 16 C_5
  + \frac{64}{3}\, C_6\right)\nonumber\\
& &
  \hspace*{-23mm}+ h(q^2,m_c) \left( \frac{4}{3}\, C_1 + C_2 + 6 C_3 + 60 C_5\right)
 -\frac{1}{2}\,h(q^2,m_b) \left( 7 C_3 + \frac{4}{3}\,C_4 + 76 C_5
  + \frac{64}{3}\, C_6\right),
\eea
and
\begin{equation}
h(q^2,m_q) = -\frac{4}{9}\, \left( \ln\,\frac{m_q^2}{\mu^2} - \frac{2}{3}
- z \right) - \frac{4}{9}\, (2+z) \sqrt{|z-1|} \times 
\left\{
\begin{array}{l@{\quad}l}
\displaystyle\arctan\, \frac{1}{\sqrt{z-1}} & z>1\\[10pt]
\displaystyle\ln\,\frac{1+\sqrt{1-z}}{\sqrt{z}} - \frac{i\pi}{2} & z \leq 1
\end{array}
\right. ,
\end{equation}
with $z=4 m_q^2/q^2$. As far as the long distance physics effects are concerned, they are encoded in two hadronic matrix elements that are conveniently expressed in terms of seven Lorentz invariant form factors. The  hadronic matrix element of the standard $V$-$A$ current is decomposed as,
\bea\label{eq:ffVA}
\langle \bar K^\ast (k) | \bar s\gamma_\mu(1-\gamma_5) b | \bar B(p)\rangle  &=&  \varepsilon_{\mu\nu\rho\sigma}\varepsilon^{*\nu} p^\rho k^\sigma\,
\frac{2V(q^2)}{m_B+m_{K^\ast}} - i \varepsilon^\ast_\mu (m_B+m_{K^\ast}) A_1(q^2)\nonumber \\
&& \hspace*{-32mm}+ i (p+k)_\mu (\varepsilon^\ast \cdot q)\, \frac{A_2(q^2)}{m_B+m_{K^\ast}}   +  i q_\mu (\varepsilon^\ast \cdot q) \,\frac{2m_{K^\ast}}{q^2}\,
\left[A_3(q^2)-A_0(q^2)\right], 
\eea
where $\varepsilon_\mu$ is the $K^\ast$-polarization vector, and thanks to the partial conservation of the axial current,
\bea
A_3(q^2) & = & \frac{m_B+m_{K^\ast}}{2m_{K^\ast}}\, A_1(q^2) -
\frac{m_B-m_{K^\ast}}{2m_{K^\ast}}\, A_2(q^2)\,,
\eea
also ensuring that no artificial divergence emerges at $q^2=0$.
Other $3$ relevant form factors parameterize the matrix element of the electromagnetic penguin operator, 
\bea\label{eq:ffT}
\langle \bar K^\ast(k) | \bar s \sigma_{\mu\nu} q^\nu (1+\gamma_5) b |
\bar B(p)\rangle &=& 2 i\varepsilon_{\mu\nu\rho\sigma} \varepsilon^{*\nu}
p^\rho k^\sigma \,  T_1(q^2)\nonumber\\
& & + \left[ \varepsilon^\ast_\mu
  (m_B^2-m_{K^\ast}^2) - (\varepsilon^\ast \cdot q) \,(2p-q)_\mu \right] T_2(q^2) \nonumber\\
& &+  
(\varepsilon^\ast \cdot q) \left[ q_\mu - \frac{q^2}{m_B^2-m_{K^\ast}^2}\, (p+k)_\mu
\right]T_3(q^2),
\eea
with $T_1(0) = T_2(0)$, ensuring that only on  form factor describes the physical $B\to K^\ast \gamma$ decay.

The explicit expression for the full differential $B\to K^\ast (\to K\pi) \ell^+\ell^-$ decay rate was presented in ref.~\cite{Kruger:1999xa}  triggered a quite intense activity in searching for  {\it good} observables, namely those that simultaneously satisfy three requirements: (i) to not suffer from large hadronic uncertainties, (ii) to have a pronounced sensitivity to the presence of physics BSM, and (iii) to be experimentally  accessible at LHCb and/or Super-B factories. Written in terms of four kinematic variables, the differential decay rate reads~\cite{Kruger:1999xa} (see also \cite{Altmannshofer:2008dz}): 
\bea\label{distr-1}
{d^4\Gamma(\bar B^0 \to \bar K^{\ast 0} \ell^+\ell^-)\over dq^2\ d\cos\theta_\ell\ d\cos\theta_K\  d\phi}={9\over 32\pi}I(q^2,\theta_\ell,\theta_K,\phi)\,,
\eea
where
\bea\label{distr-2}
I(q^2,\theta_\ell,\theta_K,\phi)&=&I_1^s(q^2) \sin^2 \theta_K + I_1^c(q^2) \cos^2\theta_K + \left[ I_2^s(q^2) \sin^2 \theta_K + I_2^c(q^2) \cos^2\theta_K\right] \cos 2\theta_\ell\nn\\
&&+ \ I_3(q^2) \sin^2 \theta_K \sin^2 \theta_\ell \cos 2\phi +  I_4(q^2) \sin 2 \theta_K \sin 2 \theta_\ell \cos \phi\nn\\
&& +\  I_5(q^2) \sin 2 \theta_K\sin\theta_\ell\cos\phi\\
&&+ \ \left[ I_6^s(q^2)\sin^2\theta_K+I_6^c(q^2)\cos^2\theta_K\right] \cos\theta_\ell +I_7(q^2) \sin 2\theta_K\sin\theta_\ell\sin\phi\nn\\
&&+\ I_8(q^2)\sin 2\theta_K\sin 2\theta_\ell\sin\phi + I_9(q^2)\sin^2\theta_K\sin^2\theta_\ell \sin 2\phi\,.\nn
\eea
Besides $q^2$, the other ki\-ne\-ma\-tical variables are defined with respect to the direction of flight of the outgoing $\bar K^{0 \ast}$  in the rest frame of $\bar B^0$. In particular  $\theta_\ell$ is the angle between that axis and lepton $\ell^-$ in the $\ell^+\ell^-$ rest frame.  $\theta_K$, instead, is the angle between that same axis and $K^-$ in the $K\pi$ rest frame. Finally $\phi$ is the angle between the $K\pi$ and ${\ell^+}{\ell^-}$ planes (see ref.~\ref{Altmannshofer:2008dz}).~\footnote{One should be careful and distinguish the CP conjugated mode $ B^0 \to K^{0 \ast}  (\to K^+\pi^-)\ell^+\ell^-$.  $\theta_\ell$ and $\phi$ are defined in the same way and are related to the ones discussed in the $ \bar B^0 \to \bar K^{0 \ast}  (\to K^-\pi^+)\ell^+\ell^-$  via $\theta_\ell \to \pi - \theta_\ell$,  $\phi \to -\phi$. The net effect on the coefficient functions in the angular distribution~(\ref{distr-2}) relevant to our discussion in this paper is that $I_6(q^2)$ and  $I_9(q^2)$ would change the sign. See refs.~\cite{Altmannshofer:2008dz,Egede:2010zc}.} The functions $I_i(q^2)$ are related to the transversity amplitudes $A_{\perp,\parallel,0,S,t}(q^2)$ as follows:
\begin{align} \label{distr-3}
I_1^s(q^2) &  =   \frac{2 + \beta^2_\ell}{4} \left[ \lvert A_{\perp}^L \rvert ^2 + \lvert A_{\parallel}^L  \rvert ^2 + \left( L \rightarrow R \right)  \right] + \frac{4m_\ell^2}{q^2} \text{Re} \left( A_{\parallel}^L  A_{\parallel}^{R \ast}  + A_{\perp}^L  A_{\perp}^{R \ast } \right)\,,\nn \\
I_1^c(q^2) & =   \lvert A_{0}^L\rvert^2 + \lvert A_{0}^R \rvert^2 + \frac{4 m_{\ell}^2}{q^2} \left[ \lvert A_t \rvert^2 +2 \text{Re} \left( A_{0}^L A_{0}^{R\ast} \right) \right] + \beta_\ell^2 \lvert A_S\rvert^2\,,\nn \\
I_2^s(q^2) & = \frac{\beta_\ell^2}{4} \left[ \lvert A_{\perp}^L \rvert ^2 + \lvert A_{\parallel}^L \rvert ^2 + \left( L \rightarrow R \right)  \right] \,,\nn\\
I_2^c(q^2) & = -\beta_\ell^2 \left( \lvert A_{0}^L\rvert^2 + \lvert A_{0}^R \rvert^2 \right)\,, \nn\\
I_3(q^2) & = \frac{1}{2} \beta_\ell^2 \left[ \lvert A_{\perp}^L\rvert ^2 -  \lvert A_{\parallel}^L \rvert ^2 +  \left( L \rightarrow R \right)  \right] \,,\nn\\
I_4(q^2) & = \frac{1}{\sqrt{2}} \beta_{\ell}^2 \left[ \text{Re} \left( A_{0}^LA_{\parallel}^{L *} \right) + \left( L \rightarrow R \right)  \right] \,,\nn \\
I_5(q^2) & = \sqrt{2} \beta_\ell \left[ \text{Re} \left( A_{0}^L A_{\perp}^{L \ast} - \left( L \rightarrow R \right)  \right) -  \frac{m_\ell}{\sqrt{q^2}}  \text{Re} \left( A_{\parallel}^L A_S^\ast + A_{\parallel}^R A_S^\ast  \right) \right]\,, \nn\\
I_6^s(q^2) & = 2 \beta_\ell \left[ \text{Re}\left( A_{\parallel}^L A_{\perp}^{L \ast} \right) - \left( L \rightarrow R \right) \right] \,, \nn\\
I_6^c(q^2) & = 4 \beta_\ell \frac{m_\ell}{\sqrt{q^2}} \left[ \text{Re}\left( A_{0}^L A_{S}^{\ast} \right) + \left( L \rightarrow R \right) \right]\,, \nn\\
I_7(q^2) & =\sqrt{2} \beta_\ell \left[ \text{Im} \left(A_{0}^L A_{\parallel}^{L  \ast} - \left( L \rightarrow R \right)  \right) + \frac{m_\ell}{\sqrt{q^2}} \text{Im} \left( A_{\perp}^L A_S^\ast+ A_{\perp}^R A_S^\ast\right) \right] \,,\nn\\
I_8(q^2) & = \frac{1}{\sqrt{2}} \beta_\ell^2 \left[ \text{Im} \left( A_{0}^LA_{\perp}^{L *} \right) +  \left( L \rightarrow R \right) \right] \,,\nn\\
I_9(q^2)   & = \beta_\ell^2\;\left[ \text{Im} \left( A_{\perp}^L A_{\parallel}^{L \ast} \right) + \left( L \rightarrow R \right) \right]\,,
\end{align}
where 
\bea
\beta_\ell =\sqrt{1-{4m_\ell^2\over q^2}}\;,
\eea
while the amplitudes $A_{\perp,\parallel,0,S,t}^{L,R}\equiv A_{\perp,\parallel,0,S,t}^{L,R}(q^2)$, when written in terms of the Wilson coefficients and the form factors, read:
\bea \label{ampl-A}
A_{\perp}^{L,R}(q^2)  &=&  \sqrt{{\cal  N} \lambda(q^2)} \bigg[ \frac{2m_b}{q^2} (C_7^\eff + C_7^{\eff\prime})\ T_1(q^2) \nn\\
&& \qquad +
\left[ (C_9^\eff + C_9^{\eff\prime}) \mp (C_{10}^\eff + C_{10}^{\eff\prime}) \right] { V(q^2) \over m_B + m_{K^\ast} }
\bigg],  \\
A_{\parallel}^{L,R}(q^2)  & =&  - \sqrt{{\cal  N}}(m_B^2 - m_{K^\ast}^2) \bigg[ 
\frac{2 m_b}{q^2} (C_7^\eff - C_7^{\eff\prime}) \ T_2(q^2)
\nn\\
&& \qquad + \left[ (C_9^\eff - C_9^{\eff\prime}) \mp (C_{10}^\eff - C_{10}^{\eff\prime}) \right] 
{A_1(q^2)\over m_B-m_{K^\ast}}
\bigg]\,, 
\eea
\begin{align}
 A_0^{L,R}(q^2) = & -\frac{\sqrt{\cal N}}{2m_{K^\ast} \sqrt{q^2}} \biggl\{  \left[ \left( C_9^{\eff} - C^{\prime\eff}_9 \right)   \mp   \left(C_{10}^{\eff} - C^{\prime \eff}_{10} \right) \right]\times \biggr.\\
		     & \qquad\left[ \left( m_B^2 -m_{K^\ast}^2 - q^2\left) \right( m_B +m_{K^\ast} \right) A_1(q^2) - \lambda(q^2)  \frac{A_2(q^2)}{m_B + m_{K^\ast}} \right] \\
		     &\qquad \left.+ 2 m_b \left(C_7^{\eff} - C^{\prime \eff}_7 \right) \left[ \left( m_B^2 + 3 m_{K^\ast}^2 - q^2\right) T_2(q_2) - \frac{\lambda(q^2)}{m_B^2 - m_{K^\ast}^2} T_3(q^2) \right] \right\}\,,\nn\\
A_t(q^2) = & \sqrt{ {\cal N} \lambda(q^2)\over q^2} \left[ 2  \left(C_{10}^{\eff} - C^{\prime\eff}_{10} \right) + \frac{q^2}{ m_\ell}  \left(C_P - C^\prime_P \right) \right]  A_0(q^2) \,, \\ 
\label{As-def} A_S(q^2) = &  -2   \sqrt{ {\cal N} \lambda(q^2)} \left( C_S - C^\prime_S \right) A_0(q^2)\,.
 \end{align}
In the above formulas,~\footnote{We chose to work in the basis of operators given in ref.~\cite{Altmannshofer:2008dz} in which  ${\mathcal{O}}_{S,P}$ and   ${\mathcal{O}}_{S,P}^\prime$  are dimension $7$ operators, and therefore the corresponding Wilson coefficients are not dimensionless but of dimension $-1$, and therefore the two terms within the brackets in $A_t(q^2)$ are of the same dimension.  }
\bea
&&{\cal N}= |V_{tb}V_{ts}^\ast|^2  \ { \beta_\ell\over 3 }\  {G_F^2 \alpha_{\rm em}^2\over  2^{10}\pi^5 m_B^3} \ q^2\ \lambda^{1/2}(q^2)\,,\nn\\
&& \nn\\
&& \lambda(q^2) = [q^2 - (m_B+m_{K^\ast})^2]\ [q^2 - (m_B-m_{K^\ast})^2]\,.
\eea

\section{Transverse asymmetries}

After integrating  eq.~(\ref{distr-1}) over $\theta_\ell$ and $\theta_K$, one arrives at:
\bea\label{eq:0}
{d^2\Gamma (B\to K^\ast\ell^+\ell^-)\over dq^2 d\phi} = {1\over 2\pi}{d\Gamma \over dq^2  }\left[ 1 + {1\over 2}  F_T(q^2)\left( A_T^{(2)}(q^2)\cos{2\phi} + A_T^{({\rm im})}(q^2)\sin{2\phi}\right)\right]\,,
\eea
where $F_T(q^2)$ is a fraction of the decay product with transversely polarized $K^\ast$, 
\bea\label{ft-def}
F_T(q^2)= \beta_\ell^2\  { \vert A_\perp(q^2)\vert^2 +\vert A_\parallel(q^2)\vert^2 \over d\Gamma/dq^2}\,.
\eea

\subsection{$A_T^{(2)}(q^2)$}
The transverse asymmetry $A_T^{(2)}(q^2)$ in eq.~(\ref{eq:0}), first introduced in refs.~\cite{Melikhov,Kruger:2005ep}, is CP-conserving and it reads
\bea\label{at2}
A_T^{(2)}(q^2)={\vert A_\perp (q^2)\vert^2 -\vert A_\parallel (q^2)\vert^2 \over \vert A_\perp (q^2)\vert^2 +\vert A_\parallel (q^2)\vert^2}\,,
\eea
where $\vert A_{\parallel,\perp} (q^2)\vert^2 = \vert A_{\parallel,\perp}^L (q^2)\vert^2 + \vert A_{\parallel,\perp}^R (q^2)\vert^2$. That quantity is expected to be experimentally accessible at LHCb and Super-B factories,  and nowadays is considered as one of the most interesting observables to study since it satisfies all three above-mentioned requirements. What is interesting to note is that $A_T^{(2)}(q^2)$ only involves $A_{\parallel,\perp} (q^2)$, and not $A_{0,S,t} (q^2)$ amplitudes. The latter ones are much more difficult to handle in QCD. In many phenomenological studies the symmetry relations among form factors, first demonstrated in ref.~\cite{Charles:1998dr}, are used to express all the form factors in terms of only two functions, $\xi_\perp(q^2)$ and $\xi_\parallel(q^2)$. The problem with that approach is that this approximation is only valid in the limit of $m_B\to \infty$ (i.e. $E_{K^\ast}\to   \infty$, at low $q^2$'s). It is not clear how to compute the non-perturbative $1/m_b^n$ corrections to that approximation, as well as those due to the finiteness of  $m_{K^\ast}$. Perturbative corrections can be handled in the QCD factorization approach~\cite{Beneke:2000wa} or via the soft collinear effective theory~\cite{Bauer:2000yr}, but we do not know how to compute the non-perturbative corrections from principles of QCD. In particular it is not clear whether or not it is even possible to formulate the problem and compute the relevant $3$-point correlation functions through numerical simulations of QCD on the lattice, and then extract the needed  $\xi_{\parallel}(q^2)$ and $\xi_{\perp}(q^2)$ form factors.  Instead, the full form factors, as defined in eqs.~(\ref{eq:ffVA},\ref{eq:ffT}), can and have been computed on the lattice.  Particularly difficult appear to be the form factors $T_3(q^2)$, $A_2(q^2)$, and  $A_0(q^2)$, that are linear combinations of $\xi_\perp(q^2)$ and $\xi_\parallel(q^2)$ as soon as the mass of $K^\ast$ is not neglected  [see eqs.~(107,109,113) in ref.~\cite{Charles:1998dr}].  On the lattice  the signals for these three form factors are particularly difficult to keep under control [see for example $A_{2,0}(q^2)$ in refs.~\cite{Abada:2002ie,Bowler:2004zb}]. The advantage of using the quantities involving only $A_{\parallel,\perp} (q^2)$ is that they do not require a detailed knowledge of these three form factors. Moreover, the ratios $A_1(q^2)/T_2(q^2)$ and $V(q^2)/T_1(q^2)$  seem to verify the symmetry relations of ref.~\cite{Charles:1998dr}  and exhibit a flat $q^2$-dependence over a wide range of $q^2$'s, satisfying the approximate relation~\footnote{ 
Since these ratios are flat for low $q^2$'s, from now on we will keep the argument of these form factors implicit. In estimating the ratios~(\ref{ratios-ff}) we take $V/T_1 \simeq 1.24$ and $A_1/T_2 \simeq 0.88$ from ref.~\cite{Ball:2004rg}, which is consistent with what is obtained through an alternative QCD sum rule method, i.e. $V/T_1 \simeq 1.22$ and $A_1/T_2 \simeq 0.92$~\cite{Colangelo:1995jv}, although the results of these two methods for the absolute values of the form factors are quite different. One of us (D.B.) checked that this feature is also verified in (quenched) lattice QCD.}  
\bea\label{ratios-ff}
{A_1/T_2 \over m_B - m_{K^\ast}} \approx {V/T_1 \over m_B + m_{K^\ast}}  \approx 0.2~\gev^{-1}\,.
\eea
Finally when replacing  $A_{\parallel,\perp} (q^2)$ in eq.~(\ref{at2}) it is convenient to factor out $T_1(q^2)$ which then cancels out in the ratio. Schematically that amounts to 
\bea
A_T^{(2)}(q^2)= {\quad \left| \ {\cal C}_\perp + {\cal C}'_\perp\ \displaystyle{{V/T_1 \over m_B + m_{K^\ast}}} \ \right| ^2 \ -\  \left(\displaystyle{T_2(q^2)\over T_1(q^2)}\right)^2 \left| \ {\cal C}_ \parallel + {\cal C}'_ \parallel \ \displaystyle{{A_1/T_2 \over m_B - m_{K^\ast}}}\  \right|^2 \quad \over 
\left| \ {\cal C}_\perp + {\cal C}'_\perp\  {\displaystyle{V/T_1 \over m_B + m_{K^\ast}}}\ \right|^2\ +\  \left(\displaystyle{T_2(q^2)\over T_1(q^2)}\right)^2 \left| \  {\cal C}_ \parallel +  {\cal C}'_ \parallel \ {\displaystyle{A_1/T_2 \over m_B - m_{K^\ast}}} \ \right|^2 }\,,
\eea
where, for notational brevity,  the combinations of Wilson coefficients and kinematical factors are denoted as $ {\cal C}^{(\prime)}_{\perp, \parallel}$. This form is useful because the ratio $T_2(q^2)/T_1(q^2)$ is a well controlled quantity and is $1$ at $q^2=0$ by definition, i.e.
\bea\label{t12}
{T_2(q^2)\over T_1(q^2)}= {T_2(0)\over T_1(0)} + z\  q^2 \equiv 1 + z\  q^2\,,
\eea
where $z$ is a slope which in a simple pole model is given by $z\approx  -1/m_{B_s^\ast}^2=-0.034~\gev^{-2}$. That value is close to $z=-0.028~\gev^{-2}$, as inferred from the light cone QCD sum rules~\cite{Ball:2004rg}, as well as to the one obtained in (quenched) lattice QCD,  $z=-0.030(3)~\gev^{-2}$~\cite{Becirevic:2006nm}. 

Therefore the advantage of using the quantities involving the amplitudes $A_{\parallel,\perp} (q^2)$, and not  $A_{0,S,t}(q^2)$, is that the relevant hadronic uncertainties are under better control. 
The price to pay when dealing only with $A_{\parallel,\perp} (q^2)$ is that no information about a coupling to the new physics scalar sector can be accessed. This, of course, can be viewed as an advantage too, because  the number of possible new physics parameters to study becomes smaller. A particular interest in studying $A_T^{(2)}(q^2)$ is the fact that its value at $q^2=0$ is identically zero in the SM, while in the extensions of SM in which the couplings to right handed currents are allowed ($\propto C_7^{\prime \eff}$) it can change drastically,   
\bea\label{eq:1}
A_T^{(2)}(0) = {2\  {\rm Re}\left[ C_7^\eff C_7^{\prime \eff \ast }\right]\over  |C_7^{\eff}|^2  + |C_7^{\prime \eff}|^2}\,.
\eea 
This argument is strictly valid only at $q^2=0$, while away from that point the situation becomes more complicated since $A_T^{(2)}(q^2)$ receives non-negligible contributions from the terms proportional to $C_{9,10}^\eff$ and  $C_{9,10}^{\prime \eff}$. 

Written in terms of $I_i(q^2)$'s explicitly given in eqs.~(\ref{distr-1},\ref{distr-2},\ref{distr-3}), $A_T^{(2)}(q^2)$ becomes particularly simple,
\bea
A_T^{(2)}(q^2) =\frac{I_3 (q^2)}{2 I_2^s (q^2)}\,.
\eea
It is important to note that there is an identity $I_1^s (q^2)= 3 I_2^s (q^2)$, valid in the case of massless leptons (reasonable assumption for $B\to K^\ast e^+e^-$). Lepton mass breaks this identity but it remains a good approximation for $q^2\gg 4 m_\ell^2$ (which in practice means a couple of $\gev^2$).

\subsection{$A_T^{({\rm im})}(q^2)$}
Measuring the term proportional to $\sin{2\phi}$ in eq.~(\ref{eq:0}) is highly interesting because the quantity 
\bea
A_T^{({\rm im})}(q^2) = - { 2 \  \text{Im} \left[ A_{\parallel}^L(q^2) A_{\perp}^{L \ast}(q^2) + A_{\parallel}^R(q^2) A_{\perp}^{R \ast}(q^2) \right]   \over \lvert A_{\bot} (q^2)\rvert^2 + \lvert A_{\parallel}(q^2) \rvert^2} \,,
\eea
is obviously identically zero in the SM for all accessible $q^2$'s. It may acquire a non-zero value only if there is a new phase, coming from physics BSM, and therefore its non-zero measurement would be a new physics discovery. In particular, at $q^2\to 0$ one obtains
\bea
 A_T^{({\rm im})}(0)=  {2\  {\rm Im}\left[ C_7^\eff C_7^{\prime \eff \ast }\right] \over  |C_7^{\eff}|^2  + |C_7^{\prime \eff}|^2 }\,,
\eea 
so that its non-zero measurement would mean that the new phase comes from the electromagnetic penguin operator, either $C_7$ or $C_7^\prime$, or both (if they are different).  As in the case of $A_T^{(2)}(q^2)$, the above formula at $q^2=0$ should be taken as asymptotic, because away from that kinematics other Wilson coefficients ($C_{9,10}^\eff$ and  $C_{9,10}^{\prime \eff}$) become important too. From the $q^2$-shape of $ A_T^{({\rm im})}(q^2)$ one can get some valuable information about the source of the new physics phase, as we will discuss in the next section of the present paper. Notice also that our definition  of $A_T^{({\rm im})}(q^2)$ is different from $A_{\rm Im}(q^2)$ introduced in ref.~\cite{Egede:2008uy}:  $2 \beta_\ell^2 A_{\rm Im}(q^2)= A_T^{({\rm im})}(q^2) \ F_T(q^2)$. 

$A_T^{({\rm im})}(q^2)$ too has a particularly simple form when written in terms of angular coefficient functions defined in eqs.~(\ref{distr-1},\ref{distr-2},\ref{distr-3}), namely,
\bea
A_T^{({\rm im})}(q^2) =\frac{I_9 (q^2)}{2 I_2^s (q^2)}\,.
\eea

\subsection{$A_T^{(5)}(q^2)$ and $A_T^{({\rm re})}(q^2)$}
Another quantity that involves only the amplitudes  $A_{\parallel,\perp} (q^2)$ has been recently proposed in ref.~\cite{Egede:2008uy}:
\bea\label{at5-def}
 A_T^{(5)}(q^2)  = { \lvert A_{\parallel}^L(q^2) A_{\perp}^{R \ast}(q^2) + A_{\perp}^L(q^2) A_{\parallel}^{R \ast}(q^2) \rvert \over \lvert A_{\bot}(q^2) \rvert^2 + \lvert A_{\parallel} (q^2)\rvert^2} \,. 
\eea
It was then shown in ref.~\cite{Egede:2010zc} that $A_T^{(5)}(q^2)$  arises naturally after realizing that eq.~(\ref{distr-1}), in the limit of massless leptons ($m_\ell=0$), respects $4$ symmetries. More specifically, the coefficient functions $I_i(q^2)$  in (\ref{distr-2}) are invariant under the following transformations:
\begin{itemize}
\item[$\circ$]  Two independent phase transformations, 
\bea
A_{i}^{L \; \prime}(q^2) =   e^{i \phi_L} A_{i}^{L}(q^2)\,, \quad  A_{i}^{R\; \prime}(q^2) =   e^{i \phi_R} A_{i}^{R}(q^2)\, \qquad (i=\perp,\parallel, 0)\,;
\eea
\item[$\circ$]  Two rotations of suitably defined vectors, $n_1= (A_\parallel^L,A_\parallel^{R\ast})$, $n_2= (A_\perp^L,-A_\perp^{R\ast})$,  $n_3= (A_0^L,A_0^{R\ast})$, 
 \bea
 R( \theta) = \left( \begin{array}{cc}  \cos  \theta  &  - \sin  \theta  \\ \sin  \theta   & \cos  \theta   \end{array} \right) \,,\qquad 
 V(\tilde{\theta}) = \left( \begin{array}{cc}  \cos \widetilde{\theta}  &  - i \sin  \widetilde{\theta}   \\ - i \sin  \widetilde{\theta}   & \cos  \widetilde{\theta}   \end{array} \right)\,.
 \eea
\end{itemize}
They eventually showed that for the massless lepton,~\footnote{Notice that in the massless case $I_1^s(q^2) = 3I_2^s(q^2)$, which is not valid if $m_\ell \neq 0$.}
\bea\label{at5-0}
A_T^{(5)}(q^2) &=&\left. {\sqrt{ 16 {I_2^s}^2(q^2) -{I_6^s}^2(q^2) -4 ( {I_3}^2(q^2) + {I_9}^2(q^2)) }  \over 8 I_2^s(q^2)}\right|_{m_\ell =0}\,.
\eea
Since we consider only the quantities constructed from $A_{\parallel,\perp}^{L,R} (q^2)$, or better the observables involving  $I_{2,6}^s(q^2)$ and $I_{3,9}(q^2)$, it suffices to consider only the first two vectors, $n_1$ and $n_2$, in which case the reasoning used in ref.~\cite{Egede:2010zc}  can be extended to the massive lepton case.  Using the two global phases, $\phi_L$ and $\phi_R$, one can first assure that  $A_{\parallel}^{L,R} (q^2)$ are real and positive. Furthermore, by choosing a suitable angle $\theta$, one can rotate $A_\parallel^L$ away and have $n_1=(0,A_\parallel^{R})$. The scalar products of vectors $n_{1,2}$ can then be easily expressed in terms of functions $I_i(q^2)$'s as follows:
\bea 
&&\lvert n_1 \rvert^2 = \left(A_{\parallel}^R(q^2)\right)^2 = {2 I_2^s(q^2) - I_3(q^2)\over  \beta_\ell^2}\,,\nn\\
 &&\lvert n_2 \rvert^2 =  \lvert A_{\perp}^L(q^2) \rvert^2 + \lvert A_{\perp}^R(q^2) \rvert^2 =   {2 I_2^s(q^2) + I_3(q^2)\over  \beta_\ell^2}\,,\nn\\ 
&& n_1 \cdot n_2       =  - A_{\parallel}^R(q^2) A_{\perp}^{R \ast} (q^2)= {\beta_\ell I_6^s(q^2) + 2 i I_9(q^2) \over 2 \beta_\ell^2}\,,
 \eea
or equivalently, 
\bea
A_{\parallel}^{L}(q^2) =  0 \,,  && A_{\parallel}^{R}(q^2) =  {1\over \beta_\ell} \sqrt{  2 I_2^s(q^2) - I_3(q^2)} \,, \nn\\
&& A_{\perp}^R (q^2)       =  - \frac{1}{2 \beta_\ell} \frac{\beta_\ell I_6^s(q^2) - 2 i I_9(q^2)}{\sqrt{2 I_2^s(q^2) - I_3(q^2) }} \,, \nn \\ 
 &&\lvert A_{\perp}^L(q^2) \rvert^2  ={1 \over 4\beta_\ell^2 } \  {16 I_2^s(q^2)^2 - 4 I_3(q^2)^2 - \beta_\ell^2 I_6^s(q^2)^2 - 4 I_9(q^2)^2 \over 2 I_2^s(q^2) -  I_3(q^2)}\,.
\eea
After inserting the last expression into eq.~(\ref{at5-def}) we then simply obtain
\bea\label{at5-i}
A_T^{(5)}(q^2) &=& {\sqrt{ 16 {I_2^s}^2(q^2) - \beta_\ell^2(q^2) {I_6^s}^2(q^2) -4 ( {I_3}^2(q^2) + {I_9}^2(q^2)) }  \over 8 I_2^s(q^2)}\,.
\eea
This formula is valid for any mass of the lepton, and obviously reproduces the massless lepton formula~(\ref{at5-0}) derived in ref.~\cite{Egede:2010zc}.  One should be very careful in constructing the observables from  $I_{i}(q^2)$, and not include those that do not respect the above symmetries.

The observable $A_T^{(5)}(q^2)$ is indeed independent from $A_T^{(2)}(q^2)$ and  $A_T^{({\rm im})}(q^2)$. In fact it is easy to see that there can be at most three independent asymmetries built up from  $A_{\perp,\parallel} (q^2)$, respecting the above symmetries.  In a generic scenario of physics BSM there can be $8$ real amplitudes: $A_{\perp}^{L,R} (q^2)$ and  $A_{\parallel}^{L,R} (q^2)$, each with its real and imaginary parts. The above four symmetries help reducing the number of independent combinations to four.~\footnote{In terms of angular coefficient functions, those four functions are: $I_2^s(q^2)$, $I_3(q^2)$, ${I_6^s}(q^2)$, and ${I_9}(q^2)$.} Since we consider the ratios, that implies the restriction to at most three such independent observables. In our case those three are: $A_T^{(2)}(q^2)$,  $A_T^{({\rm im})}(q^2)$, and $A_T^{(5)}(q^2)$. 

From eq.~(\ref{at5-i}) it is clear that $A_T^{(5)}(q^2)$ is too complicated a quantity. After introducing  $A_T^{(2)}(q^2)$ and  $A_T^{({\rm im})}(q^2)$, the only new angular coefficient function needed for $A_T^{(5)}(q^2)$ is  $I_6^s(q^2)$. We find therefore more convenient to introduce,
\bea
 A_T^{({\rm re})}(q^2) =  \frac{ 2 \  \text{Re} \left[ A_{\parallel}^L(q^2) A_{\perp}^{L \ast}(q^2) - A_{\parallel}^R(q^2) A_{\perp}^{R \ast}(q^2) \right] }{\lvert A_{\bot} (q^2)\rvert^2 + \lvert A_{\parallel}(q^2) \rvert^2}  =  \frac{\beta_\ell \ I_6^s(q^2)}{4\ I_2^s(q^2)}\,.
\eea
$A_T^{(5)}(q^2)$ is related to $ A_T^{(2)}(q^2)$, $ A_T^{({\rm im})}(q^2)$ and  $A_T^{({\rm re})}(q^2)$ via
\begin{equation}
 \left( 2 A_T^{(5)}(q^2)\right)^2 +  \left( A_T^{(2)}(q^2) \right)^2 +  \left( A_T^{({\rm im})}(q^2)\right)^2 +  \left(  A_T^{({\rm re}) }(q^2)\right)^2  = 1 \,.
\end{equation}
It is worth mentioning that in the scenarios in which the new physics does not modify the scalar amplitude, $A_S(q^2)$ [c.f. eq.~(\ref{As-def})], the angular coefficient function $I_6^c(q^2)=0$, and $A_T^{({\rm re})}(q^2)$ is  related to the usual  forward-backward asymmetry as 
\bea
A_T^{{\rm (re)}}(q^2) = \frac{4\beta_\ell}{3} \ \frac{A_{\rm FB}(q^2) }{F_T (q^2)}\,,
\eea
where  $F_T (q^2)= 4 I_2^s(q^2)/(d\Gamma/dq^2)$ is the quantity we already defined in eq.~(\ref{ft-def}), whereas 
\bea
A_{\rm FB}(q^2) = {\displaystyle{
\int_0^1 d\cos\theta_\ell  {d^2\Gamma \over dq^2 d\cos\theta_\ell} \ -\  \int_{-1}^0 d\cos\theta_\ell {d^2\Gamma \over dq^2 d\cos\theta_\ell} } \over \displaystyle{d\Gamma/dq^2}} = {3\over 8} \ {  2 I_6^s(q^2)+I_6^c(q^2)\over \displaystyle{d\Gamma/dq^2}}\,.
\eea
The expected shape of  $A_{\rm FB}(q^2)$ in the SM and its various extensions  has been discussed in great detail in the literature~\cite{FB}. Importantly, the property that ${\displaystyle\lim_{q^2\to 0}}A_{\rm FB}(q^2)\to 0$, caries over to $A_T^{{\rm (re)}}(q^2)$. Therefore, unlike $ A_T^{(2)}(0)$ and $A_T^{{\rm (im)}}(0)$ whose values at $q^2=0$  can change considerably if the new physics affects the Wilson coefficients $C_7^{(\prime)}$, the third quantity  $A_T^{{\rm (re)}}(0)$ remains insensitive to new physics. However the $q^2$-shapes of our three asymmetries can teach us about new physics. In particular the point at which these three quantities might become zero can be helpful in discerning among BSM scenarios.

\subsection{How to extract $A_T^{({\rm re})}(q^2)$ from experiment?}

Before passing onto the potentially interesting phenomenology that one can learn from the above asymmetries at low $q^2$'s, we need to discuss the possibility to experimentally measure the above introduced asymmetry $A_T^{{\rm (re)}}(q^2)$ [and therefore $A_T^{(5)}(q^2)$ too].  Since $A_T^{({\rm re})}(q^2)$ resembles the standard forward-backward asymmetry, a way to extract it from experiment is quite similar to that employed for  $A_{\rm FB}(q^2)$. 
After integrating eq.~(\ref{distr-1}) in $\phi$, one can separate the events with the lepton going forward from those in which lepton flies backwards and get the differential distribution of the forward-backward asymmetry, 
\bea
{d^2 A_{\rm FB}(q^2,\theta_K)\over dq^2 d\cos\theta_K} &=&{  \displaystyle{ \left[ \int_0^1 - \int_{-1}^0 \right] }d\cos\theta_\ell \int_0^{2\pi} d\phi {\displaystyle{ d^4\Gamma(B\to K^\ast \ell^+\ell^-)\over dq^2\ d\cos\theta_\ell\ d\cos\theta_K\  d\phi}} 
\over  \displaystyle{d\Gamma/dq^2}} \nn\\
&&\cr
&=& {9\over 16 \displaystyle{(d\Gamma/dq^2)}} \left[  I_6^s(q^2) \sin^2\theta_K   +  I_6^c(q^2) \cos^2\theta_K \right] \nn\\
&=& F_T(q^2) {9\over 16\beta_\ell } \left[   A_T^{({\rm re})}(q^2) \sin^2\theta_K + \beta_\ell {I_6^c(q^2)\over 4 I_2^s(q^2) } \cos^2\theta_K   \right] \,.
\eea
As before, $F_T(q^2)$ is the quantity specified in eq.~(\ref{ft-def}), and  $A_T^{({\rm re})}(q^2)$ can be simply read off from the angular dependence of the forward-backward asymmetry. 
The term proportional to $\cos^2\theta_K$ is expected to be small: in fact in the SM  $I_6^c(q^2)$ is exactly zero, while in the BSM scenarios it can become important only if the new scalar contributions are large.~\footnote{From the experimental upper bound $B(B_s\to \mu^+\mu^-)$~\cite{PDG,exp-bsmu} one can constrain $|C_S|< 0.8/m_b \lesssim 0.2~\gev^{-1}$.} However, even in that latter case $I_6^c(q^2)$ remains negligibly small if one considers $B\to K^\ast e^+e^-$,  because the scalar amplitude $A_S(q^2)$  enters in $I_6^c(q^2)$ multiplied by the lepton mass. In the $B\to K^\ast \mu^+\mu^-$ mode, instead, one should check whether or not the term  $\propto \cos^2\theta_K$ is discernible.

\section{What can we learn from the shapes of  $A_T^{(2),({\rm re}),({\rm im})}(q^2)$ and from $A_T^{(2),({\rm re}),({\rm im})}(q_0^2)\to 0$?}

\begin{figure}[t!]
\psfrag{aa}{\color{blue}\Huge $q^2$} 
\psfrag{Standard}{\color{blue}\huge \hspace*{-30mm}$ $} 
\begin{center}
{\resizebox{11cm}{!}{\includegraphics{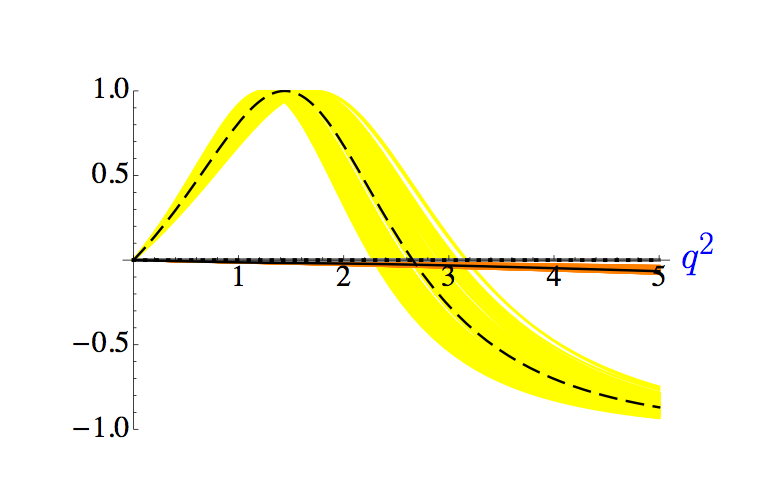}}} 
\caption{\label{fig:St}\footnotesize{\sl 
Transverse asymmetries in the Standard Model ($C_7^\prime=C_9^\prime=C_{10}^\prime=0$). Note that $A_T^{{\rm (im)}}(q^2)$ (dotted line) is identically zero, $A_T^{{\rm (2)}}(q^2)$ (full line) indistinguishable from zero for low values of $q^2$'s, while $A_T^{{\rm (re)}}(q^2)$ varies between $1$ and $-1$.  }} 
\end{center}
\end{figure}

As we already mentioned, the quantities $A_T^{{\rm (im)}}(q^2)$ and  $A_T^{{(2)}}(q^2)$ can considerably change the value at $q^2=0$ in the presence of new physics contributions.  $A_T^{{\rm (re)}}(q^2)$ instead scales as $q^2$ at low $q^2$'s, and is zero at $q^2=0$ in the SM and its extensions. Even the functional dependence of $A_T^{{\rm (re)}}(q^2)$ is not too sensitive to the presence of new physics, which is different from what happens with $A_T^{{\rm (im)}}(q^2)$ and  $A_T^{{(2)}}(q^2)$. It is interesting to examine the situations in which these asymmetries can change the sign, while remaining in the low $q^2$-region.

To measure a non-zero value for the asymmetry  $A_T^{{\rm (im)}}(q^2)$, a new phase is needed. The origin of that phase can be figured out from the functional dependence  of $A_T^{{\rm (im)}}(q^2)$ at low $q^2$'s.  In particular $A_T^{{\rm (im)}}(q^2)$ can go through zero at $q^2_{0 \ {\rm (im)}}$ if there is a solution to the equation:
\bea\label{satim}
&&{\rm Im}\left\{ C_7^{\eff\prime}C_7^{\eff\ast} - {1\over 2} {q^2\over 2 m_b} \left[ (C_7^{\eff} - C_7^{\eff\prime}) (C_9^{\eff}+C_9^{\eff\prime})^\ast \ {V(q^2)/T_1(q^2) \over m_B + m_{K^\ast}}\right.\right.\nn\\
&&\left. \hspace*{42mm}+ (C_7^{\eff} + C_7^{\eff\prime})^\ast (C_9^{\eff}-C_9^{\eff\prime}) \ {A_1(q^2)/T_2(q^2) \over m_B - m_{K^\ast}} \right]\nn\\
&&\hspace*{7mm} \left.+\left({q^2\over 2m_b}\right)^2 (  C_9^{\eff\prime}C_9^{\eff\ast}  + C_{10}^{\eff\prime}C_{10}^{\eff\ast} ) {V(q^2)/T_1(q^2) \over m_B + m_{K^\ast}}  {A_1(q^2)/T_2(q^2) \over m_B - m_{K^\ast}} \right\}=0\,.
\eea

On the other hand, measuring the point at which $A_T^{{\rm (re)}}(q^2)$ crosses the $q^2$-axis resembles the standard discussion of $A_{\rm FB}(q_0^2)=0$, and the corresponding $q^2_{0 \ {\rm (re)}}$ is defined via
\begin{align}\label{satre}
&{\rm Re}\left\{ {q^2\over m_b^2} ( C_{10}^{\eff }C_9^{\eff \ast} - C_{10}^{\eff\prime}C_9^{\eff\prime \ast} ) \ {A_1(q^2)/T_2(q^2) \over m_B - m_{K^\ast}} \  {V(q^2)/T_1(q^2) \over m_B + m_{K^\ast}}  \right.\nn\\
& \hspace*{11mm}  + (C_7^{\eff} -  C_7^{\eff\prime})  (C_{10}^{\eff} + C_{10}^{\eff\prime})^\ast   {V(q^2)/T_1(q^2) \over m_B + m_{K^\ast}} \nn\\
& \hspace*{11mm}+ \left.   (C_7^{\eff} +  C_7^{\eff\prime})^\ast  (C_{10}^{\eff} - C_{10}^{\eff\prime}) \  {A_1(q^2)/T_2(q^2) \over m_B - m_{K^\ast}}  \right\}=0\,.
\end{align}

A similar formula for determining $q^2_{0 \ {\rm (2)}}$ at which  $A_T^{{(2)}}(q^2)$ crosses zero cannot be written in a simple form. To find a simple formula, it suffices to note that for small $q^2$'s a following approximation is reasonable, $T_1(q^2) \lambda^{1/2}(q^2) \approx T_2(q^2) (m_B^2-m_{K^\ast}^2)$, so that $A_T^{{(2)}}(q^2) = 0$ can be written as 
\begin{align}\label{sat2}
 {q^2\over m_b^2}   {\rm Re}& \left[ (C_7^{\eff} + C_7^{\eff\prime}) (C_9^{\eff}+C_9^{\eff\prime})^\ast {V(q^2)/T_1(q^2) \over m_B + m_{K^\ast}} -  (C_7^{\eff} - C_7^{\eff\prime})^\ast (C_9^{\eff}-C_9^{\eff\prime}) {A_1(q^2)/T_2(q^2) \over m_B - m_{K^\ast}} \right] \nn \\
&+  \left( {q^2\over 2 m_b^2}\right)^2 \left[ \left( {V(q^2)/T_1(q^2) \over m_B + m_{K^\ast}} \right)^2 [|C_9^{\eff}+C_9^{\eff\prime}|^2 + |C_{10}^{\eff} + C_{10}^{\eff\prime}|^2]\right.\nn\\
& \left. - \left( {A_1(q^2)/T_2(q^2) \over m_B - m_{K^\ast}} \right)^2 [|C_9^{\eff}-C_9^{\eff\prime}|^2 + |C_{10}^{\eff} - C_{10}^{\eff\prime}|^2]\right] +  4 {\rm Re}(C_7^{\eff} C_7^{\eff\prime\ast}) =0\,.
\end{align}
Furthermore it is reasonable to consider $C_9^\eff(q^2)$ to be $q^2$-independent at low $q^2$'s, because in that region the function $Y(q^2)$ in eq.~(\ref{eq:Y}) varies very slowly at low $q^2$-region. In fact, $Y(q^2)$ makes about $20\%$ of the whole $C_9^\eff(q^2)$, so that its small variation in $q^2$ might at worst entail an error of a couple of percents.  In the following we will therefore consider $Y(q^2)=1$.  
To situate the positions of $q^2_{0 \ {\rm (im)}}$, $q^2_{0 \ {\rm (re)}}$ and $q^2_{0 \ {(2)}}$ it is helpful to use eq.~(\ref{ratios-ff}), that we will denote by $R$. The expressions for $q^2_0$~(\ref{satim},\ref{satre},\ref{sat2})  simplify, but their solutions are not very compact unless we restrain our attention to the specific NP scenarios. We should emphasize once again that the discussion above refers to the case in which $C_7$ remains negative, i.e. of the same sign as in the SM. If that was not the case, $A_T^{{\rm (re)}}(q^2)$ would have not crossed the $q^2$-axis.

\subsection{New physics in $C_7$ and $C_7^\prime$}
From now on, for notational simplicity, we will drop the superscript ``eff" from the Wilson coefficients. If $C_9$ and $C_{10}$ are kept fixed to their SM values, and $C_9^\prime  = C_{10}^\prime=0$, the points at which 
the asymmetries  $A_T^{{(2)}}(q^2)$, $A_T^{{\rm (im)}}(q^2)$ and $A_T^{{\rm (re)}}(q^2)$ may change the sign are obtained by solving eqs.~(\ref{sat2},\ref{satim}, \ref{satre}) respectively. We obtain:
\bea
q^2_{0 \ {(2)}} &=& -{2 m_b\over R} {\ {\rm Re} \left( C_7^\ast C_7^\prime\right)\ \over C_9 {\rm Re}\left(  C_7^\prime\right)}\,,\cr
&&\cr
q^2_{0 \ {\rm (im)}} &=& -{2 m_b\over R} {\ {\rm Im} \left( C_7^\ast C_7^\prime\right)\ \over C_9 {\rm Im}\left(  C_7^\prime\right)}\,,\cr
&&\cr
q^2_{0 \ {\rm (re)}} &=& -{2 m_b\over R} {\ {\rm Re} \left( C_7\right)\ \over C_9 }\,.
 \eea
where $C_7, C_7^\prime\in \mathbb{C}$ is understood. At this point we should reiterate that the approximations discussed above~\footnote{For reader's commodity we repeat that our approximations at low $q^2$'s are: (1)$(A_1/T_2)/(m_B - m_{K^\ast})   = (V/T_1)/(m_B + m_{K^\ast})=R$, (2) $\partial C_9^\eff/\partial q^2 = 0$, and $Y(q^2)=1$, and (3) in the case of  $A_T^{{\rm (2)}}(q^2)$ we also use $T_1(q^2) \lambda^{1/2}(q^2) = T_2(q^2) (m_B^2-m_{K^\ast}^2)$.  }  are used to obtain the simple expressions for $q^2_{0 \ {\rm (im)}}$, $q^2_{0 \ {\rm (re)}}$ and $q^2_{0 \ {(2)}}$, but all the plots presented in this work are obtained by using full expressions. Excellent agreement between the approximate and full results confirms the validity of our approximations. 
\begin{figure}[t!!]
\begin{center}
{\includegraphics[scale=0.41]{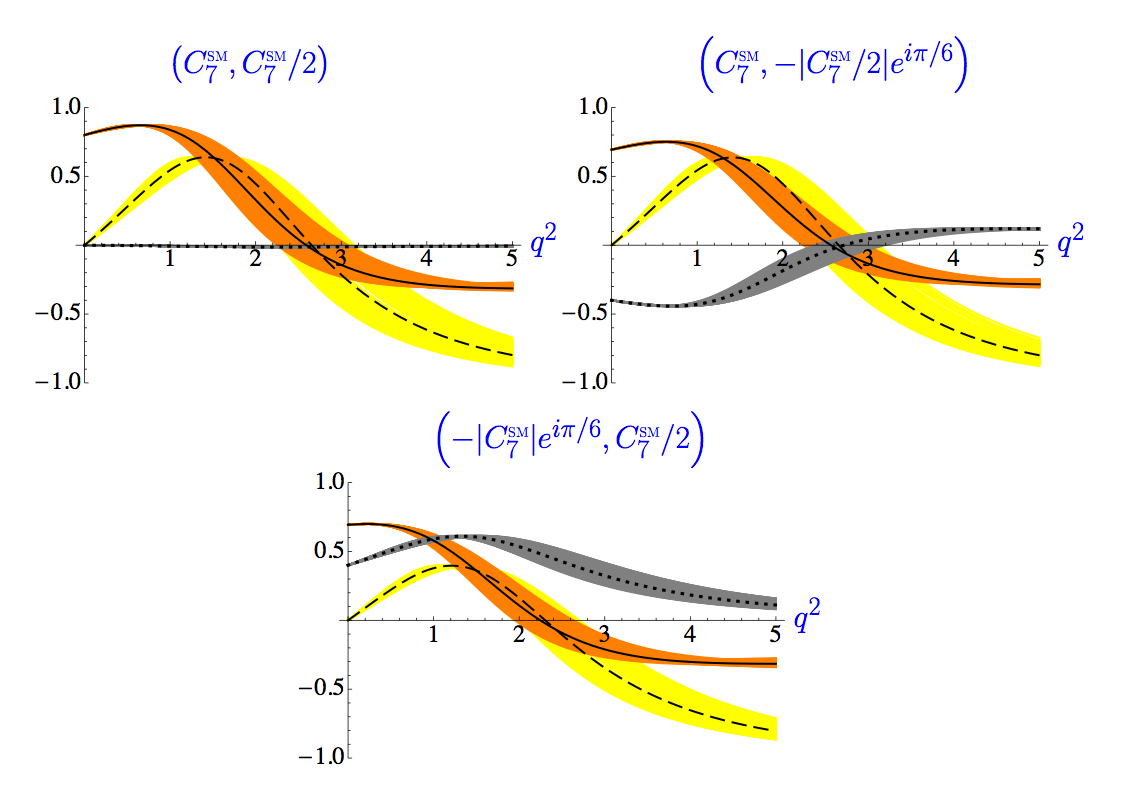}}
\caption{\label{fig:7}\footnotesize{\sl 
Dependence of the asymmetries  $A_T^{{\rm (2)}}(q^2)$ (full line), $A_T^{{\rm (im)}}(q^2)$ (dotted line),  $A_T^{{\rm (re)}}(q^2)$ (dashed line) on $q^2$ for different possibilities (real or complex) of  $(C_7,C_7^\prime)$, while $(C_9,C_9^\prime)$ and  $(C_{10},C_{10}^\prime)$ are kept fixed to their SM values.  }} 
\end{center}
\end{figure}
One can distinguish three interesting situations:
\begin{itemize}
\item For $C_7, C_7^\prime\in \mathbb{R} $, and $ C_7^\prime\neq 0$, one has 
\bea
q^2_{0 \ {(2)}} =q^2_{0 \ {\rm (re)}} = -{2 m_b\over R} {C_7\over C_9}\,,\quad\quad q^2_{0 \ {\rm (im)}} -\text{does not exist}\,.
\eea
If $ C_7^\prime= 0$, the situation becomes the SM-like, namely $q^2_{0 \ {\rm (re)}}$ remains the same, while $q^2_{0 \ {(2)}}$ does not exist anymore (see also fig.~\ref{fig:St}).
\item Particularly interesting is the situation with $C_7 \in \mathbb{R} $ and $C_7^\prime\in \mathbb{C} $, in which all three asymmetries cross the $q^2$-axis at the same point, 
\bea
q^2_{0 \ {(2)}} =q^2_{0 \ {\rm (im)}}= q^2_{0 \ {\rm (re)}} = -{2 m_b\over R} {C_7\over C_9}\,.
\eea
Of course this coincidence of zeroes could be spoiled if our approximations were bad. We checked that three asymmetries indeed become zero at almost the same point even when the full expressions are used, as shown in fig.~\ref{fig:7}.
\item In a reverse situation, i.e. $C_7 \in \mathbb{C} $ and $C_7^\prime\in \mathbb{R}$, one has
\bea
q^2_{0 \ {(2)}} =q^2_{0 \ {\rm (re)}} = -{2 m_b\over R} { {\rm Re} (C_7)\over C_9}\,,\quad \quad q^2_{0 \ {\rm (im)}} -\text{does not exist}\,.
\eea
\end{itemize} 
It is important to note that the above discussion holds in the situations in which the new physics does not alter the sign of the Wilson coefficient $C_7$. If that happens then obviously the asymmetries do not have zeroes anymore.~\footnote{That issue has not been resolved by the recent results on $A_{\rm FB}(q^2)$ measured at BaBar and Belle~\cite{FB-exp}.} We stress again that in order to produce the plots in fig.~\ref{fig:7} we did not use any approximation to the full formula.  To illustrate the situations in which the new phase alter the SM values of the Wilson coefficients, we choose that phase to be $\pi/6$, but without altering the sign of ${\rm Re}(C_7^{(\prime)})$ with respect to the SM value. In other words, we take for $C_7 \in \mathbb{C}$, $C_7 = - \vert C_7^{\rm SM}\vert \displaystyle{e^{i\pi/6}} = C_7^{\rm SM} \displaystyle{e^{i\pi/6}}$. Similarly, when $C_7^\prime \in \mathbb{C}$, the illustrations are provided by using $C_7 = - \vert C_7^{\rm SM}\vert/2 \displaystyle{e^{i\pi/6}}$.  The bands of the curves shown in all the plots of this paper are obtained from Monte Carlo in which we {\it uniformly} varied the form factor ratios $R$, the slope $z$, and the quark masses within the ranges specified in Appendix.

\subsection{New physics in $C_9$ and $C_9^\prime$}
\begin{figure}[h!!]
\begin{center}
{\includegraphics[scale=0.42]{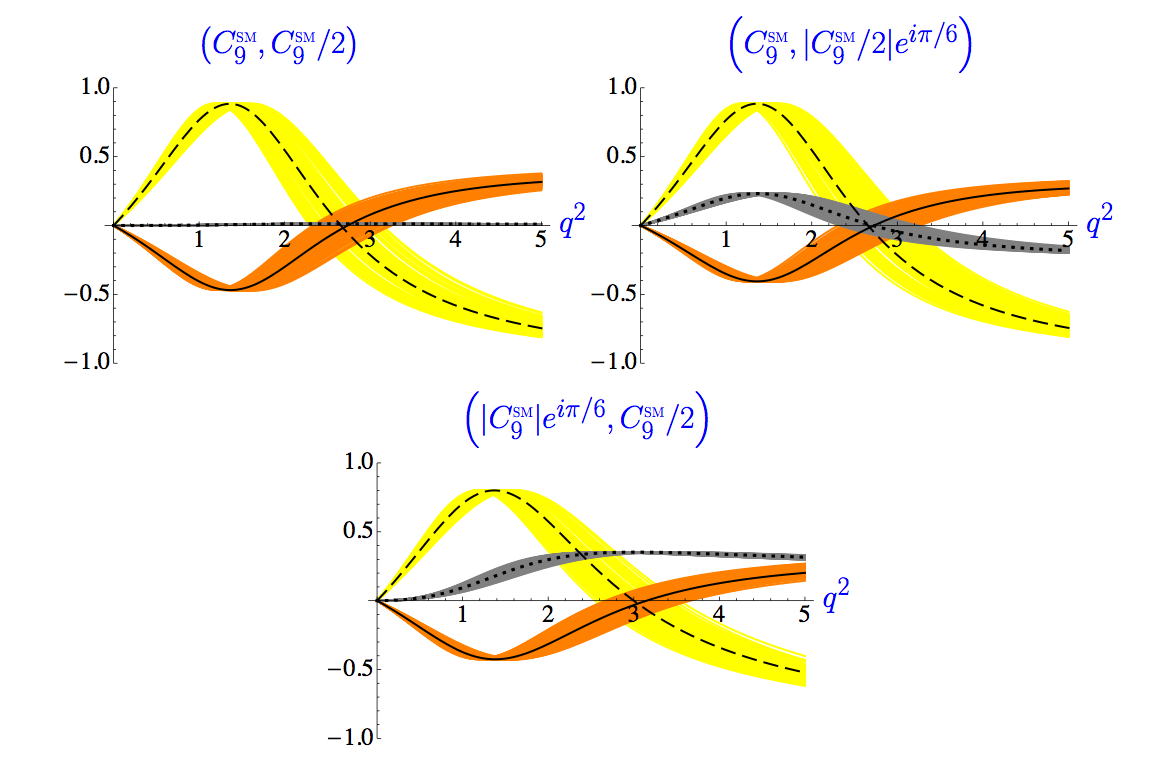}}
\caption{\label{fig:9}\footnotesize{\sl 
Similar to fig.~\ref{fig:7}, plotted is the $q^2$-dependence of three asymmetries  $A_T^{{\rm (2)}}(q^2)$ (full line), $A_T^{{\rm (im)}}(q^2)$ (dotted line),  $A_T^{{\rm (re)}}(q^2)$ (dashed line), for different possibilities of new physics in  $(C_9,C_9^\prime)$ (see the text), while $(C_7,C_7^\prime)$ and  $(C_{10},C_{10}^\prime)$ are kept fixed at their SM values.  }} 
\end{center}
\end{figure}
Another distinct possibility is to consider the case in which  $C_9$ and $C_9^\prime$ are affected by the new physics effects, while $C_7$ and $C_{10}$  remain intact (and $C_7^\prime=C_{10}^\prime =0$). The points $q^2_{0 \ (2),{\rm (im),(re)}}$, where the asymmetries have zeroes, are: 
\bea
q^2_{0 \ {(2)}} &=& -{2 m_b\over R} {\ C_7 \ {\rm Re}\ \left(  C_9^\prime\right)\ \over  {\rm Re}\left(  C_9 C_9^\prime\right)}\,,\cr
&&\cr
q^2_{0 \ {\rm (im)}} &=& -{2 m_b\over R} {\  C_7\  {\rm Im}\ \left(  C_9^\prime\right)\ \over {\rm Im}\left(  C_9  C_9^\prime\right)}\,,\cr
&&\cr
q^2_{0 \ {\rm (re)}} &=& -{2 m_b\over R} {\ C_7 \ \over {\rm Re}\left( C_9\right) }\,.
 \eea
Three interesting situations in this case are:
\begin{itemize}
\item If there is no new phase, i.e. $C_9, C_9^\prime\in \mathbb{R} $, one has~\footnote{At $q^2\geq 4m_c^2$,   ${\rm Im}[Y(q^2)]\neq 0$, and therefore $C_9$ becomes complex. That region is however above the point at which the asymmetries cross the $q^2$-axis. } 
\bea
q^2_{0 \ {(2)}} =q^2_{0 \ {\rm (re)}} = -{2 m_b\over R} {C_7\over C_9}\,,\quad \quad q^2_{0 \ {\rm (im)}} -\text{does not exist}\,.
\eea
\item For $C_9 \in \mathbb{R} $ and $C_9^\prime\in \mathbb{C} $, the three zeroes again coincide, 
\bea
q^2_{0 \ {(2)}} =q^2_{0 \ {\rm (im)}}= q^2_{0 \ {\rm (re)}} = -{2 m_b\over R} {C_7\over C_9}\,.
\eea
\item If, instead, $C_9 \in \mathbb{C} $ and $C_9^\prime\in \mathbb{R}$, then
\bea
q^2_{0 \ {(2)}} =q^2_{0 \ {\rm (re)}} = -{2 m_b\over R} { C_7\over {\rm Re} (C_9)}\,,\quad \quad q^2_{0 \ {\rm (im)}} = 0\,.
\eea
\end{itemize} 

These three situations are illustrated in fig.~\ref{fig:9}. We see that, like in fig.~\ref{fig:7}, when one of the two Wilson coefficients is complex while the other one is real, two or three asymmetries cross the $q^2$-axis at the same point. That information alone would not tell whether the new physics contribution modifies $C_7^{(\prime)}$ or $C_9^{(\prime)}$. However the low $q^2$'s shapes of the asymmetries $A_T^{{\rm (2)}}(q^2)$ and $A_T^{{\rm (im)}}(q^2)$ in fig.~\ref{fig:9} are very different from those in fig.~\ref{fig:7}, which solves the ambiguity.

\subsection{New physics in $C_{10}$ and $C_{10}^\prime$}
\begin{figure}[ht!!]
\begin{center}
{\includegraphics[scale=.42]{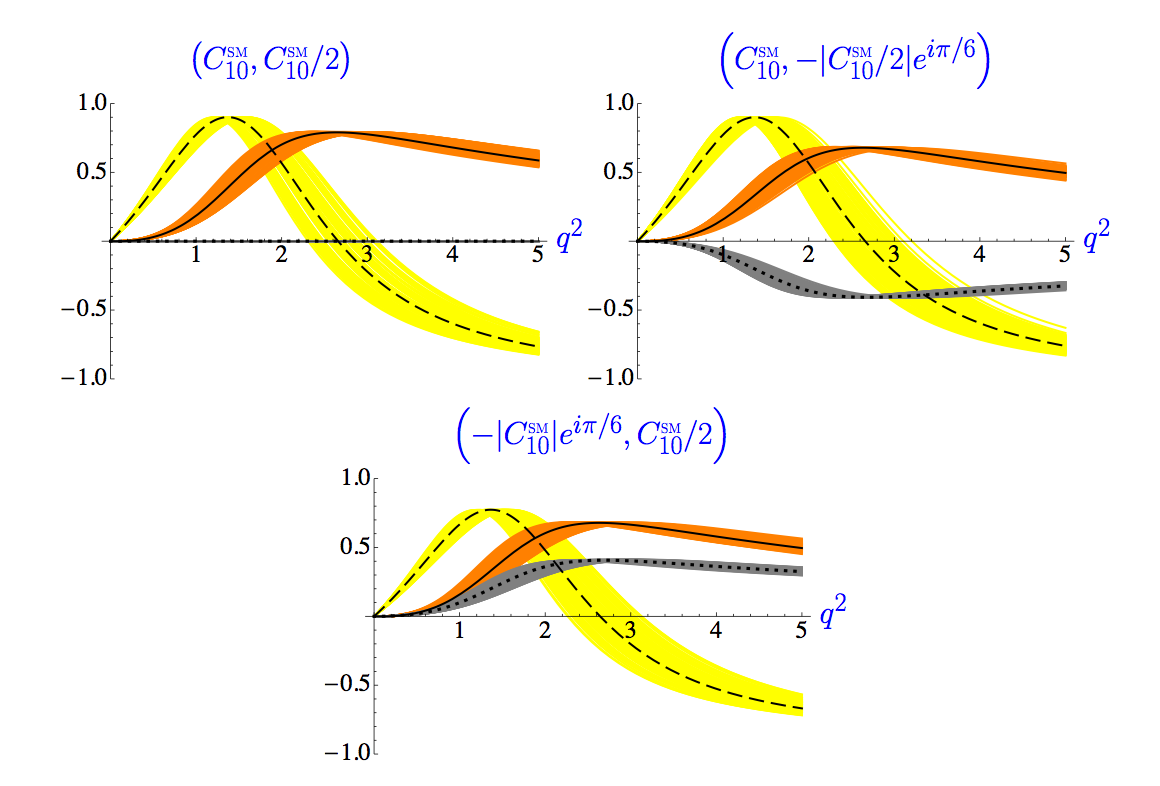}}
\caption{\label{fig:10}\footnotesize{\sl 
Expected $q^2$-shapes of asymmetries $A_T^{{\rm (2)}}(q^2)$ (full line), $A_T^{{\rm (im)}}(q^2)$ (dotted line),  $A_T^{{\rm (re)}}(q^2)$ (dashed line) , if the new physics alters only  $(C_{10},C_{10}^\prime)$, and $(C_7,C_7^\prime)$ and  $(C_{9},C_{9}^\prime)$ remain at their SM values.  }} 
\end{center}
\end{figure}
The third special case we consider here is the one in which $C_{10}$ and $C_{10}^\prime$ receive non-zero contributions from new physics, while $C_7$ and $C_9$ retain their SM values. 
The zeroes of our three asymmetries then become:
\bea
q^2_{0 \ {(2)}} =  q^2_{0 \ {\rm (im)}} = 0,\quad  q^2_{0 \ {\rm (re)}} = -{2 m_b\over R} {\ C_7 \ \over  C_9  }\,,
 \eea
independent on $C_{10}$ and $C_{10}^\prime$. 

In this case, neither $A_T^{{\rm (2)}}(q^2)$  nor $A_T^{{\rm (im)}}(q^2)$ crosses the $q^2$-axis away from zero. The $q^2$ dependencies of the three asymmetries are different from those found in the previous two cases, as it can be appreciated by comparing fig.~\ref{fig:10} with those presented in the previous two subsections. 

\subsection{Maximum of $A_T^{{\rm (re)}}(q^2)$}

In all this discussion we supposed the new physics would not change the sign of the real part of the WIlson coefficients $C_{7,9}$, i.e. their signs in the SM.  From the plots presented in figs.~\ref{fig:St}, \ref{fig:7}, \ref{fig:9} and \ref{fig:10}, we see that the shape of $A_T^{{\rm (re)}}(q^2)$ remains stable under the variation of Wilson coefficients, even in the presence of a new physics phase. Its maximal value can provide us with another interesting information. Using the approximations discussed above,  in the SM, we get that the point $\widetilde q^2$ at which $dA_T^{{\rm (re)}}(q^2)/dq^2=0$ is 
\bea
\widetilde q^2 = {2 m_b\over R}\ {C_7\over C_{10}-C_9} \approx 1.3~\gev^2\quad \Longrightarrow \quad A_T^{{\rm (re)}}(\widetilde q^2) = 1\,.
\eea
We checked numerically that, even without using our approximations, this result remains valid. In the presence of new physics, the value of $\widetilde q^2$ only slightly shifts  with respect to the SM one, but {\underline{$A_T^{{\rm (re)}}(\widetilde q^2)$ becomes lower}}. The only exception to this rule is the situation in which the right-handed currents are absent and all the Wilson coefficients receive the same phase. 
Concerning the three new physics scenarios discussed in this section, we note that   $A_T^{{\rm (re)}}(\widetilde q^2)$ is considerably lower in the case of new physics modifying the $(C_7,C_7^\prime)$ values: for   $(C_7,C_7^\prime)=( C_7^{\rm SM}, \displaystyle{e^{i \varphi}}  |C_7^{\rm SM}/2|)$, one gets $A_T^{{\rm (re)}}(\widetilde q^2)\approx 0.6$ regardless of the value of the phase $\varphi$;  for  $(C_7,C_7^\prime)=( -|C_7^{\rm SM}| \displaystyle{e^{i \varphi}},   C_7^{\rm SM}/2)$, the suppression of this asymmetry is even stronger but depends on  $\varphi$  [in particular, for $\varphi=\pi/6$ we have $A_T^{{\rm (re)}}(\widetilde q^2)\approx 0.4$].

\section{Summary}
In this paper we discussed the benefits of using three asymmetries,  $A_T^{{\rm (2)}}(q^2)$, $A_T^{{\rm (im)}}(q^2)$ and  $A_T^{{\rm (re)}}(q^2)$, built up from  the amplitudes  $A_{\perp} (q^2)$ and  $A_{\parallel} (q^2)$  and interesting for studying the presence of physics BSM in the $B\to K^\ast \ell^+\ell^-$ decay. In fact this is the maximal number of independent  asymmetries  invariant under the transformations discussed in ref.~\cite{Egede:2010zc}. Involving only  $A_{\perp} (q^2)$ and  $A_{\parallel} (q^2)$, these asymmetries are less prone to hadronic uncertainties. We show that a study of their low $q^2$-dependence ($q^2 < m_{J/\psi}^2$) can help discerning among various new physics scenarios.~\footnote{$A_T^{{\rm (2)}}(q^2)$ and other transverse asymmetries have been discussed in the literature~\cite{Altmannshofer:2008dz,Kruger:2005ep,Egede:2010zc,transv-more}. Here we focus onto those that we consider phenomenologically more interesting because their smaller sensitivity to hadronic uncertainties.} In particular from the shapes of $A_T^{{\rm (2)}}(q^2)$ and $A_T^{{\rm (im)}}(q^2)$ at low $q^2$'s and the point at which the three asymmetries can change the sign, one can tell which operators receive  contributions from physics BSM.   

The asymmetry  $A_T^{{\rm (im)}}(q^2)$ has a pleasant property that it is exactly zero in the SM, and in the presence of a new physics phase its value can become significantly different from zero, and therefore its experimental study would be highly welcome. 

Furthermore, we introduced $A_T^{{\rm (re)}}(q^2)$, which is a simpler quantity than the commonly used $A_T^{{\rm (5)}}(q^2)$, and it is related to it via 
\bea \left[ 2 A_T^{(5)}(q^2)\right]^2 + \left[  A_T^{(2)}(q^2)\right]^2 +  \left[  A_T^{({\rm im})}(q^2)\right]^2 + \left[   A_T^{({\rm re}) }(q^2)\right]^2  = 1\,.\nn
\eea
Since our quantities involve only  $A_{\perp,\parallel} (q^2)$, we could extend the expression for $A_T^{{\rm (5)}}(q^2)$, derived in ref.~\cite{Egede:2010zc},  and write $A_T^{{\rm (5)}}(q^2)$ in terms of functions $I_i(q^2)$ entering the angular distributions of  $B\to K^\ast \ell^+\ell^-$ to the massive lepton case as well.  If the new physics does not alter the sign of ${\rm Re} \ C_{7}$ or ${\rm Re} \ C_{9}$, we note that our asymmetry $A_T^{{\rm (re)}}(q^2)$ reaches the maximum at low $q^2$ and is equal to one in the SM, while it gets suppressed in the presence of right handed currents and/or the new physics phase. 

Measuring all three asymmetries discussed in this paper should be within reach at LHCb and the Super-B factories. 
\vspace{3 cm}

\section*{Acknowledgments}
It is our pleasure to thank J.~LeFran\c{c}ois and M.-H.~Schune for discussions, J.~Matias for his valuable comments on the manuscript, J.~Virto and A.~Tayduganov for pointing two important typos in the previous version of this paper. The support of the French ANR via the contract ÓLFV-CPV-LHCÓ ANR-NT09-508531 is kindly acknowledged. The work by E.~Schneider is helped in part by LLP/Azione Erasmus.
\newpage

\section*{Appendix}
In the numerical analysis we used $m_B=5.279$~GeV, and $m_{K^\ast}=0.892$~GeV. For the SM Wilson coefficients we take~\cite{Altmannshofer:2008dz}:
\bea
C_1=-0.257,\quad C_2=1.009,\quad  C_3=-0.005 ,\quad C_4=-0.078,\quad C_5=0.000\nn\\
C_6=0.001,\quad C_7^{\rm eff}=-0.304,\quad  C_9^{\rm eff}-Y(q^2)=4.211 ,\quad C_{10}^{\rm eff}=-4.103,
\eea
computed to next-to-next-to leading logarithmic accuracy in $\msbar$(NDR) renormalization scheme at the scale $\mu=4.8~\gev$~\cite{Bobeth:1999mk}. 
For the $b$ and the charm quark mass we take~\cite{PDG}:
\bea
m_c^\msbar (m_c)=1.28(8)~\gev\,,\qquad
m_b^\msbar (m_b)=4.19\left({}_{-06}^{+18}\right)~\gev\,.
\eea
 Concerning the form factors in the range of low $q^2$'s, the ratio $R$ and the slope $z$, defined in eqs.~(\ref{ratios-ff}) and ~(\ref{t12}) respectively, have been uniformly (not Gaussianly) varied within the ranges~\cite{Ball:2004rg,Colangelo:1995jv,Becirevic:2006nm}:
\bea
R\in  (0.17, 0.23)~\gev^{-1}\,,\qquad z \in - (0.027,0.033)~\gev^{-2}\,. 
\eea


\vspace{3 cm}

\begin{thebibliography}{99}

\bibitem{Heff}
  B.~Grinstein, M.~J.~Savage and M.~B.~Wise,
  Nucl.\ Phys.\  B {\bf 319} (1989) 271;
  M.~Misiak,
  Nucl.\ Phys.\  B {\bf 393} (1993) 23
  [Erratum-ibid.\  B {\bf 439} (1995) 461];
A.~J.~Buras and M.~Munz,
  Phys.\ Rev.\  D {\bf 52} (1995) 186
  [arXiv:hep-ph/9501281];



\bibitem{Bobeth:1999mk}
  C.~Bobeth, M.~Misiak and J.~Urban,
  Nucl.\ Phys.\  B {\bf 574} (2000) 291
  [arXiv:hep-ph/9910220].



\bibitem{Altmannshofer:2008dz}
  W.~Altmannshofer, P.~Ball, A.~Bharucha, A.~J.~Buras, D.~M.~Straub and M.~Wick,
  JHEP {\bf 0901} (2009) 019
  [arXiv:0811.1214 [hep-ph]].


\bibitem{Wilsoneff}
A.~J.~Buras, M.~Misiak, M.~M\"unz and S.~Pokorski,
  Nucl.\ Phys.\  B {\bf 424} (1994) 374
  [arXiv:hep-ph/9311345].

\bibitem{Melikhov}
D.~Melikhov, N.~Nikitin and S.~Simula,
  Phys.\ Lett.\  B {\bf 442} (1998) 381
  [arXiv:hep-ph/9807464].

\bibitem{Kruger:1999xa}
  F.~Kruger, L.~M.~Sehgal, N.~Sinha and R.~Sinha,
  Phys.\ Rev.\  D {\bf 61} (2000) 114028
  [Erratum-ibid.\  D {\bf 63} (2001) 019901]
  [arXiv:hep-ph/9907386].


\bibitem{Kruger:2005ep}
  F.~Kruger and J.~Matias,
  Phys.\ Rev.\  D {\bf 71} (2005) 094009
  [arXiv:hep-ph/0502060].



\bibitem{Charles:1998dr}
  J.~Charles, A.~Le Yaouanc, L.~Oliver, O.~Pene and J.~C.~Raynal,
  Phys.\ Rev.\  D {\bf 60} (1999) 014001
  [arXiv:hep-ph/9812358].



\bibitem{Beneke:2000wa}
  M.~Beneke and T.~Feldmann,
  Nucl.\ Phys.\  B {\bf 592} (2001) 3
  [arXiv:hep-ph/0008255].

\bibitem{Bauer:2000yr}
  C.~W.~Bauer, S.~Fleming, D.~Pirjol and I.~W.~Stewart,
  Phys.\ Rev.\  D {\bf 63} (2001) 114020
  [arXiv:hep-ph/0011336].

\bibitem{Abada:2002ie}
  A.~Abada, D.~Becirevic, P.~Boucaud, J.~M.~Flynn, J.~P.~Leroy, V.~Lubicz and F.~Mescia
                  [SPQcdR collaboration],
  Nucl.\ Phys.\ Proc.\ Suppl.\  {\bf 119} (2003) 625
  [arXiv:hep-lat/0209116].


\bibitem{Bowler:2004zb}
  K.~C.~Bowler, J.~F.~Gill, C.~M.~Maynard and J.~M.~Flynn  [UKQCD
                  Collaboration],
  JHEP {\bf 0405} (2004) 035
  [arXiv:hep-lat/0402023].



\bibitem{Ball:2004rg}
  P.~Ball and R.~Zwicky,
  Phys.\ Rev.\  D {\bf 71} (2005) 014029
  [arXiv:hep-ph/0412079].


\bibitem{Colangelo:1995jv}
  P.~Colangelo, F.~De Fazio, P.~Santorelli and E.~Scrimieri,
  Phys.\ Rev.\  D {\bf 53} (1996) 3672
  [Erratum-ibid.\  D {\bf 57} (1998) 3186]
  [arXiv:hep-ph/9510403].



\bibitem{Becirevic:2006nm}
  D.~Becirevic, V.~Lubicz and F.~Mescia,
  Nucl.\ Phys.\  B {\bf 769} (2007) 31
  [arXiv:hep-ph/0611295].



\bibitem{Egede:2008uy}
  U.~Egede, T.~Hurth, J.~Matias, M.~Ramon and W.~Reece,
  JHEP {\bf 0811} (2008) 032
  [arXiv:0807.2589 [hep-ph]].

\bibitem{Egede:2010zc}
  U.~Egede, T.~Hurth, J.~Matias, M.~Ramon and W.~Reece,
  JHEP {\bf 1010} (2010) 056
  [arXiv:1005.0571 []].




\bibitem{FB}
A.~Bharucha and W.~Reece,
  Eur.\ Phys.\ J.\  C {\bf 69} (2010) 623
  [arXiv:1002.4310 [hep-ph]].
  A.~K.~Alok, A.~Dighe, D.~Ghosh, D.~London, J.~Matias, M.~Nagashima and A.~Szynkman,
  JHEP {\bf 1002} (2010) 053
  [arXiv:0912.1382 [hep-ph]];
A.~Hovhannisyan, W.~S.~Hou and N.~Mahajan,
  Phys.\ Rev.\  D {\bf 77} (2008) 014016
  [arXiv:hep-ph/0701046];
P.~Colangelo, F.~De Fazio, R.~Ferrandes and T.~N.~Pham,
  Phys.\ Rev.\  D {\bf 73} (2006) 115006
  [arXiv:hep-ph/0604029];
T.~Feldmann and J.~Matias,
  JHEP {\bf 0301} (2003) 074
  [arXiv:hep-ph/0212158];
A.~Ali, P.~Ball, L.~T.~Handoko and G.~Hiller,
  Phys.\ Rev.\  D {\bf 61} (2000) 074024
  [arXiv:hep-ph/9910221];
  T.~M.~Aliev, C.~S.~Kim and Y.~G.~Kim,
  Phys.\ Rev.\  D {\bf 62} (2000) 014026
  [arXiv:hep-ph/9910501]; 
 D.~Melikhov, N.~Nikitin and S.~Simula,
  Phys.\ Lett.\  B {\bf 442} (1998) 381
  [arXiv:hep-ph/9807464];
A.~Ali, T.~Mannel and T.~Morozumi,
  Phys.\ Lett.\  B {\bf 273} (1991) 505;


\bibitem{PDG}
  K.~Nakamura {\it et al.}  [Particle Data Group],
  ``Review of particle physics,''
  J.\ Phys.\ G {\bf 37} (2010) 075021.

\bibitem{exp-bsmu}
  R.~Aaij {\it et al.}  [the LHCb Collaboration],
  Phys.\ Lett.\  B {\bf 699} (2011) 330
  [arXiv:1103.2465 [hep-ex]].


\bibitem{FB-exp}
B.~Aubert {\it et al.}  [BABAR Collaboration],
  Phys.\ Rev.\  D {\bf 79} (2009) 031102
  [arXiv:0804.4412 [hep-ex]];
  J.~T.~Wei {\it et al.}  [BELLE Collaboration],
  Phys.\ Rev.\ Lett.\  {\bf 103} (2009) 171801
  [arXiv:0904.0770 [hep-ex]].



\bibitem{transv-more}
C.~Bobeth, G.~Hiller and D.~van Dyk,
  JHEP {\bf 1007} (2010) 098
  [arXiv:1006.5013 [hep-ph]], 
also  c.f.  arXiv:1105.2659 [hep-ph];
  E.~Lunghi and J.~Matias,
  JHEP {\bf 0704} (2007) 058
  [arXiv:hep-ph/0612166];
S.~Descotes-Genon, D.~Ghosh, J.~Matias and M.~Ramon,
  arXiv:1104.3342 [hep-ph];
A.~K.~Alok, A.~Datta, A.~Dighe, M.~Duraisamy, D.~Ghosh, D.~London and S.~U.~Sankar,
  arXiv:1008.2367 [hep-ph].




\end{thebibliography}
\end{document}